\documentstyle[11pt,emulateapj]{article} 

\begin{document}

\pagestyle{myheadings}
\markboth{Tanuma, S., Yokoyama, T., Kudo, T., \& Shibata, K. 2000, ApJ}
{Supernova-Shock-Driven Magnetic Reconnection}
%
%
%

\title{
Two-Dimensional MHD Numerical Simulations of\\
Magnetic Reconnection Triggered by\\
A Supernova Shock in Interstellar Medium:\\
Generation of X-Ray Gas in Galaxy}

\author{\center
Syuniti Tanuma \altaffilmark{1,2,3},
Takaaki Yokoyama, \altaffilmark{4},
Takahiro Kudoh \altaffilmark{2,5},
\and
Kazunari Shibata \altaffilmark{6}}

\altaffiltext{1}{
Department of Astronomy, School of Science, University of Tokyo,
7-3-1 Hongo, Bunkyo-ku, Tokyo 113-0033, Japan}
\altaffiltext{2}{National Astronomical Observatory of Japan,
2-21-1 Osawa, Mitaka-shi, Tokyo 181-8588, Japan}
\altaffiltext{3}{Present Address:
Solar-Terrestrial Environment Laboratory, Nagoya University,
3-13 Honohara, Toyokawa, Aichi 442-8507, Japan
: tanuma@stelab.nagoya-u.ac.jp} 
\altaffiltext{4}{Nobeyama Radio Observatory,
Minamimaki, Minamisaku, Nagano 384-1305, Japan
: yokoyama@nro.nao.ac.jp}
\altaffiltext{5}{kudoh.takahiro@nao.ac.jp}
\altaffiltext{6}{
Kwazan Observatory, Faculty of Science, Kyoto University,
Ominemachi, Kitakazan, Yamashina-ku, Kyoto 607-8471, Japan
: shibata@kwasan.kyoto-u.ac.jp}

\tighten

\received{5, Mar, 2000}
\accepted{??, Sep, 2000}
\journalid{100}{??, ???, 2001}
\articleid{10}{20}
\paperid{AAS}
\ccc{1}
\cpright{AAS}{2000}



\begin{abstract}

We examine the magnetic reconnection triggered by a supernova
(or a point explosion) in interstellar medium,
by performing two-dimensional resistive magnetohydrodynamic (MHD)
numerical simulations with high spatial resolution.
We found that
the magnetic reconnection starts long after a supernova shock
(fast-mode MHD shock) passes a current sheet.
The current sheet evolves as follows:
(i) Tearing-mode instability is excited by the supernova shock,
and the current sheet becomes thin in its nonlinear stage.
(ii) The current-sheet thinning is saturated
when the current-sheet thickness becomes comparable to
that of Sweet-Parker current sheet.
After that, Sweet-Parker type reconnection starts,
and the current-sheet length increases.
(iii) ``Secondary tearing-mode instability'' occurs 
in the thin Sweet-Parker current sheet.
(iv) As a result, further current-sheet thinning occurs
and anomalous resistivity sets in, 
because gas density decreases in the current sheet.
Petschek type reconnection starts and heats interstellar gas.
Magnetic energy is released quickly while magnetic islands are moving
in the current sheet during Petschek type reconnection.
The released magnetic energy
is determined by the interstellar magnetic field strength,
not energy of initial explosion nor distance to explosion.
We suggest that magnetic reconnection is a possible mechanism
to generate X-ray gas in Galaxy.
\\
\keywords{
magnetic reconnection ---
Galaxy: structure --- ISM: magnetic fields ---
MHD --- shock waves --- X-rays: ISM}

\end{abstract}

\section{INTRODUCTION}


Observations of the diffuse component of X-rays (e.g. Snowden 1995)
indicate that hot components of temperatures higher than $10^6$ K
occupy a considerable volume of the interstellar space.
These hot components may be generated by supernova remnants (SNRs),
stellar winds and magnetic heating.
X-rays from Galactic ridge (GRXE; Galactic Ridge X-ray Emission;
see discussion; \cite{kan97}; \cite{koy86a}; \cite{mak94};
\cite{yam97}; \cite{yam96}; \cite{val98}),
for example, can not be fully explained
by the point sources such as SNRs and stellar winds (\cite{kan97}).
We study the magnetic heating in interstellar space in this paper.
The mean strength of interstellar magnetic field is several $\mu$G
(\cite{bec96}; \cite{hei76}; \cite{sof86}; \cite{val97}).
The interstellar magnetic field has random components
and generates possibly many current sheets at all scales
(e.g. \cite{ohn93}).
Parker (1992) pointed out the importance of magnetic reconnection 
for the heating of Galactic plasma (see also \cite{han98}).
Magnetic reconnection is a fundamental intrinsic property of 
agitated magnetized, turbulent plasma.
Whenever the magnetic fields collide with
another field with different direction by, for example, the shear motion,
magnetic energy is rapidly dissipated.
Outside the forming current sheet
the magnetic field lines are frozen-in to the interstellar gas,
which is provided by the high electrical conductivity.
When the oppositely directed field lines collide with each other,
the field gradient steepens and the current density
increases until strong dissipation sets in 
(e.g. via anomalous resistivity)
to trigger fast reconnection (e.g. \cite{han98}; \cite{uga86}).
The reconnection heats the interstellar gas 
by releasing interstellar magnetic energy,
and accelerates it by magnetic tension force to Alfv\'en velocity,
forming interstellar jets.
The magnetic reconnection plays an important role 
as observed in solar flares by X-ray satellite
{\it Yohkoh} (\cite{tsu96}; \cite{shi96}),
and in substorms by satellite such as {\it Geotail} (e.g. \cite{hos98}).
In the solar atmosphere, the magnetic reconnection heats
the plasma from a temperature of several $\times 10^6$ K
to several $\times 10^7$ K (sometimes several $\times 10^8$ K),
and accelerates it to $10^{2-3}$ km s$^{-1}$ 
(e.g. \cite{shi96}; \cite{yok95}).
If an explosion such as a supernova compresses the current sheet,
magnetic reconnection occurs and heats interstellar gas.

Historically, the steady reconnection mechanisms were proposed
on the basis of analytical studies.
In the Sweet(1958)-Parker(1957) type reconnection,
the diffusion region is so long as to occupy whole current system.
Therefore, the stored magnetic energy is released by Ohmic heating,
which can hardly be applied to solar flare phenomena,
because the reconnection rate of the Sweet-Parker model is too small
($\sim R_{\rm m}^{-1/2}$) in the solar corona,
where $R_{\rm m}$ is magnetic Reynolds number ($\sim 10^{10-15}$).
On the other hand, in the Petschek(1964) type reconnection,
the diffusion region is localized near an X-point,
and standing slow shocks occupy whole current systems.
In the configuration, the motor effect associated with slow shocks
is much more dominant than Ohmic heating,
which can hence be applicable to solar flare phenomena,
because the reconnection rate of the Petschek model is $\sim 0.1-0.01$,
independent of $R_{\rm m}$ when $R_{\rm m}$ is large.
This is called ``fast reconnection''.
Two models have been suggested to clarify 
how the fast reconnection can be realized.
One is so-called ``externally driven fast reconnection model'',
which predicts that the fast reconnection should be controlled 
by boundary conditions (\cite{pet64}; \cite{sat79}; \cite{pri86}).
The other is so-called ``spontaneous fast reconnection model'',
which predicts that the fast reconnection mechanism spontaneously
develops from inside the system by self-consistent
interaction between plasma microscopic processes
(current-driven anomalous resistivities) and
macroscopic reconnection flows (see also \cite{uga86}).

The purpose of this paper is to examine the magnetic reconnection
triggered by a supernova by performing two dimensional (2D)
numerical magnetohydrodynamic (MHD) simulations
including the electric resistivity.
Many MHD numerical simulations have been performed
for magnetic reconnection in the solar atmosphere
(e.g. \cite{yok95}; \cite{ods97}).
Some similar attempts have been made in the model of intracluster medium
(\cite{val96}; see also \cite{taj97}),
Galactic halo (\cite{bir98}; \cite{tan98}; \cite{zim97}),
and protostar (\cite{hay96}; \cite{shi99}).
The magnetic reconnection occurs at numerous current sheets
in Galactic disk and halo,
which may be called ``galactic flare'' (\cite{stu80}).
Recently, Tanuma et al.\ (1999) performed numerical simulations
and suggested that the magnetic reconnection can generate
the Galactic-ridge X-ray emission (see section \ref{grxe}).
In this paper,
we solve the 2D resistive MHD equations numerically
to see how the magnetic reconnection occurs
when a supernova triggers the magnetic reconnection in Galaxy.
In this paper,
we analize the detailed time evolution of the reconnecting current sheet,
study dependence of the results on many parameters,
and apply the results to Galaxy.

In the next section,
we describe the method of the 2D resistive MHD numerical simulations
of the magnetic reconnection triggered by a supernova
as the simplest model.
In section 3, we describe the results of the simulations,
and we apply the results to the X-ray plasmas in Galaxy in section 4.
In the last section, we summarize this paper.

\section{METHOD OF NUMERICAL SIMULATION} 
 
\subsection{The Situation of Problem}

The magnetic reconnection occurs
when the interstellar magnetic field collides with another magnetic field
which is not exactly parallel.
The magnetic reconnection heats and accelerates interstellar gas
by releasing magnetic energy.
In the present paper, we treat the magnetic reconnection 
triggered by a supernova (figure \ref{Situation}).
We assume a simple initial condition (uniform temperature everywhere,
and uniform magnetic field outside the current-sheet)
and large simulation region.
We assume the magnetic fields are anti-parallel to each other,
and a supernova occurs near the current sheet.
We solved the two-dimensional (2D) resistive magnetohydrodynamic (MHD)
equations numerically 
to see how the magnetic reconnection heats the interstellar gas.

\subsection{MHD Basic Equations}

The 2D resistive MHD basic equations are written as follows:
\begin{eqnarray}
{\partial\rho\over\partial t}+\nabla\cdot(\rho\mbox{\boldmath$v$})&=&0,\\
\rho{\partial\mbox{\boldmath$v$}\over\partial t}
+\rho(\mbox{\boldmath$v$}\cdot\nabla)\mbox{\boldmath$v$}
+\nabla p_{\rm g}
&=& {1\over c} \mbox{\boldmath$J$}\times\mbox{\boldmath$B$}
+\rho\mbox{\boldmath$g$},\\
{\partial\mbox{\boldmath$B$}\over\partial t}
-\nabla\times(\mbox{\boldmath$v$}\times\mbox{\boldmath$B$})
&=& -c\nabla\times(\eta\mbox{\boldmath$J$}),\\
{\partial e\over\partial t}
+\nabla\cdot\left[(e+p_{\rm g})\mbox{\boldmath$v$}\right]
&=& \eta |\mbox{\boldmath$J$}|^2
+\mbox{\boldmath$v$}\cdot\nabla p_{\rm g}
\end{eqnarray}
where $\rho$, $\mbox{\boldmath$v$}$, $\mbox{\boldmath$B$}$,
$\eta$, $e$, $\mbox{\boldmath$J$}$, and $\mbox{\boldmath$g$}$
are mass density, velocity, magnetic field,
electric resistivity, internal energy,
current density ($=c\nabla\times\mbox{\boldmath$B$}/4\pi$),
and gravity (=0), respectively.
We use the equation of state for the ideal gas,
i.e., $p_{\rm g}=(\gamma-1)e$
where $\gamma$ is the specific heat ratio (=5/3).

\subsection{Normalization \label{Normalization}}

We normalize length, velocity, and time. The units are
\begin{eqnarray}
H &\simeq& 100E_{51}^{1/3}
\left({p_{13}\over 3}\right)^{-1/3}\ \rm pc,
\label{H}\\
C_{\rm s}
&=& (\gamma R_{\rm g}T_0)^{1/2}
\simeq 20T_4^{1/2}\ \rm km\ s^{-1},
\label{Cs}\\
\tau &=& {H\over C_{\rm s}}
\simeq
3\times 10^6E_{51}^{1/3}T_4^{-1/2}
\left({p_{13}\over 3}\right)^{-1/3}\ \rm yr,
\label{tau}
\end{eqnarray}
respectively,
where $E_{51}$, $T_4$, $p_{13}$, $C_{\rm s}$, and $R_{\rm g}$
are the initial explosion energy ($E_{\rm ex}$) in unit of $10^{51}$ erg,
temperature in unit of $10^4$ K,
gas pressure in units of $10^{-13}$ erg cm$^{-3}$,
sound velocity, and gas constant, respectively.
We assume that $H$ is derived from
\begin{equation}
E_{\rm ex}={p_{\rm ex}\over\gamma-1}
           \Bigl({4\pi\over 3}H^2D_{\rm SN}\Bigr).
\label{Esn}
\end{equation} 
where $D_{\rm SN}$ $(=H)$ is a half-thickness of
assumed initial high-pressure region
in a direction perpendicular to $x$-$y$ plane,
and $p_{\rm ex}=p_{\rm ratio}p_{\rm g}=500p_{\rm g}
=5\times 10^{-11}p_{13}$ (Model A1).

The units of density, gas pressure, magnetic field strength, 
current density, and resistivity are 
$\rho_0\sim 1\times 10^{-25}$ g cm$^{-3}$,
$\rho_0 C_{\rm s}^2\sim 4\times 10^{-13}$ erg cm$^{-3}$,
$(\rho_0 C_{\rm s}^2/8\pi)^{1/2}\sim 3.2\ \mu$G,
$(\rho_0 C_{\rm s}^2/8\pi)^{1/2}/H$,
and $c^2HC_{\rm s}/4\pi$, respectively.
In this paper, we solve non-dimensional basic equations,
and describe the results of numerical simulations
by non-dimensional values.

\subsection{Anomalous Resistivity Model}

We assume the anomalous resistivity model as follows:
\begin{equation}
\eta=\left\{\begin{array}{ll}
\eta_0                 & \mbox{if}\ v_{\rm d}\leq v_{\rm c} \\
\eta_0+\alpha (v_{\rm d}/v_{\rm c}-1)^2
                      & \mbox{if}\ v_{\rm d}  >  v_{\rm c}
            \end{array}\right.
\label{eta}
\end{equation}
(\cite{uga92}; \cite{yok95}),
where $v_{\rm d}$($\equiv J/\rho$), $\rho$, $J$, $v_{\rm c}$, and $\eta_0$
are the normalized relative ion-electron drift velocity,
non-dimensional mass density, non-dimensional current density,
threshold above which anomalous resistivity sets in,
and ``background uniform resistivity''.
The parameters are $\eta_0=0.015$,
%
%
$\alpha=0.1$ and $v_{\rm c}=100$ in this paper
for the typical model (Model A1).

\subsection{Initial Condition of Typical Model (Model A1)}

Figure \ref{Init} shows the initial condition of our numerical simulation.
We assume the temperature is $T=T_0=1$ (uniform) everywhere.
The current-sheet thickness is $2l^{\rm init}=2$.
We take Cartesian coordinate ($x,y$).
The magnetic field, gas pressure, and density are
\begin{eqnarray}
\mbox{\boldmath$B$}&=&B_0\tanh (y)\mbox{\boldmath$x$},\\
p_{\rm g}&=&p_0+p_{\rm mag}^{\rm init}\cosh^{-2}(y),\\
\rho&=&\gamma p_{\rm g}/T
=\rho_0+(\gamma p_{\rm mag}^{\rm init}/T_0)\cosh^{-2}(y),
\end{eqnarray}
respectively
where $B_0\simeq 8.68$, $\mbox{\boldmath$x$}=(1,0)$,
$p_0=1/\gamma=0.6$, $\rho_0=1$,
and $p_{\rm mag}^{\rm init}$ is initial magnetic pressure
($=B_0^2/8\pi=p_0/\beta$) outside the current sheet.
The ratio of gas to magnetic pressure is
$\beta=8\pi p_0/B_0^2=0.2$ ($|y|>1$).
Total pressure ($P=p_{\rm g}+|\mbox{\boldmath$B$}|^2/8\pi$)
is uniform everywhere.
The sound velocity is $C_{\rm s}\equiv (\gamma p_{\rm g}/\rho)^{1/2}=1$
(uniform) in the initial condition.
The initial Alfv\'en velocity is 
$v_{\rm A}^{\rm init}=B_0/(4\pi\rho_0)^{1/2}\simeq 2.45$ ($|y|>1$).

We assume the symmetric boundaries for the top ($y\simeq 81.7$)
and bottom ($y\simeq -54.8$) surfaces,
and the periodic boundaries for the left ($x\simeq -120.9$)
and right ($x\simeq 120.9$) surfaces.
The simulation region size is
$L_{\rm x}\sim 241.8$ times $L_{\rm y}\sim 136.5$.
The number of grid points is
($N_{\rm x}\times N_{\rm y})=(465\times 602)$.
The grid sizes are non-uniform
($\triangle x\geq 0.20$, $\triangle y\geq 0.025$).
The number of grid points in the current sheet is $80$.
We put a high-pressure region instead of a point explosion (supernova)
at ($x,y$)=($0,y_{\rm ex}$)=($0,7$),
whose radius and gas pressure are $r_0=1$ and
$p_{\rm ex}=p_{\rm ratio}p_{\rm g}=500p_0$, respectively.

We use the 2-steps modified Lax-Wendroff method.
We neglect gravitational force,
effects of rotation such as shearing motion and Coriolis force,
cosmic rays, viscosity, radiative cooling, and heat conduction
in our simulation models.
The cooling times due to heat conduction and radiation are
\begin{eqnarray}
\tau_{\rm cond}
&\sim& {nkT\lambda^2\over\kappa_0 T^{7/2}}\nonumber\\
&\sim& 10^{13}({n\over 0.1\ {\rm cm}^{-3}})\nonumber\\
& &({\lambda\over 100\ {\rm pc}})^2
({T\over 10^4\ {\rm K}})^{-5/2}\ {\rm yr},\\
\tau_{\rm rad}
&\sim& {nkT\over n^2\Lambda(T)}\nonumber\\
&\sim& 10^3({T\over 10^4\ {\rm K}})
({n\over 0.1\ {\rm cm}^{-3}})^{-1}\nonumber\\
& &\left[{\Lambda(10^4\ {\rm K})
   \over 10^{-21}\ {\rm erg\ cm^3\ s^{-1}}}\right]^{-1}\ {\rm yr},
\end{eqnarray}
respectively, for cool, dense gas,
where $\Lambda$ is the cooling function (\cite{spi62}),
and $\kappa_0$ is constant ($=10^{-6}$ erg s$^{-1}$ cm$^{-1}$ K$^{-1}$).
%
The radiative cooling time becomes 
$\tau_{\rm rad}\sim 10^6(n/0.1\ {\rm cm}^{-3})^{-1}$ yr
for the hot gas of $T\sim 10^7$ K,
where $\Lambda(10^7\ {\rm K})$ is $\sim 10^{-23}$ erg cm$^3$ s$^{-1}$.
Furthermore,
in the X-ray gas near Galactic plane (Galactic Ridge X-ray Emission=GRXE;
$T\sim 10^8$ K and $n\sim 3\times 10^{-3}$ cm$^{-3}$; see Discussion),
the radiative cooling time is $\sim 10^{10.5}$ yr,
which is much longer than the typical time scale
($3\times 10^6$ yr; see next section)
of the physical process examined in this paper,
so that the radiative cooling can be neglected.
%
On the other hand, the conduction cooling time for the X-ray gas
is $\sim 30$ yr [which is
$\sim 10^5(\lambda_{\rm eff}/1\ {\rm kpc})(T/10^8\ {\rm K})^{-1/2}$ yr
in fact
because GRXE gas is collisionless] $\ll 3\times 10^6$ yr,
so that the conduction cooling can not be neglected,
where $\lambda_{\rm eff}$ is the effective length of helical magnetic field.
Nevertheless, we neglect the effect of heat conduction
for simplicity in this paper,
since the basic properties of magnetic reconnection
such as reconnection rate nor energy-release rate are not much
affected by the heat conduction (\cite{yok97}).
We assume so long current sheet that 
the effect of evaporation is negligible,
though the evaporation occurs 
if the interstellar cloud exists near the reconnection region
and the reconnected field lines penetrate the cloud.
In this paper,
we assume the uniform strong field and long current sheet
under pressure equilibrium for simplicity,
instead of a small turbulence or small current sheets
under non-equilibrium.

\section{RESULTS}
\subsection{Typical Model (Model A1)}

Figure \ref{InitDe} shows the time evolution of the two-dimensional
distribution of the current density, density, and gas pressure,
with magnetic field lines,
in an early phase of the interaction between
the supernova and current sheet.
A supernova occurs near the current sheet.
A blast shock passes and perturbs the current sheet
(figure \ref{InitDe}).
Figures \ref{Temperature}-\ref{Current}
show the time evolution of spatial distributions of
the temperature (figure \ref{Temperature}),
gas pressure (figure \ref{Pressure}),
and current density (figure \ref{Current}),
with velocity vectors and magnetic field lines.
Figure \ref{Scenario} shows the schematic illustration of simulation results.
The current sheet evolves and magnetic reconnection occurs
sequentially as the following 5 phases.
(i) Shock wave passage ($t<2$; figure \ref{Scenario}[a])
and the current sheet thinning by the tearing instability in nonlinear phase
($t\sim 100-200$; figure \ref{Scenario}[b]).
(ii) Sweet-Parker type reconnection
($t\sim 200-270$; figure \ref{Scenario}[c]).
(iii) ``Secondary tearing instability'' in the Sweet-Parker current sheet
($t\sim 250-270$; figure \ref{Scenario}[d]).
(iv) Generation of magnetic islands
($t\sim 270$; figure \ref{Scenario}[e])
and Petschek type (fast) reconnection
with two standing slow shock regions
($t\sim 270-400$; figure \ref{Scenario}[f]).
(v) End of Petschek type reconnection ($t\sim 400$).

The current sheet is unstable to the tearing mode instability (\cite{fur63}).
The magnetic dissipation time, Alfv\'en time,
and time scale of the tearing instability 
in the initial state are
\begin{eqnarray}
\tau_{\rm dis}^{\rm init}
&=& {l^{\rm init}{}^2\over\eta_0}
\sim {1^2\over 0.015}
\sim 67,\\
\tau_{\rm A}^{\rm init}
&=& {l^{\rm init}\over v_{\rm A}^{\rm init}}
\sim {1\over 2.5}
\sim 0.4,\\
\tau_{\rm t}^{\rm init}
&=& (\tau_{\rm dis}^{\rm init}\tau_{\rm A}^{\rm init})^{1/2}
\sim 5.2,
\end{eqnarray}
respectively,
where $l^{\rm init}(=1)$ is the half-thickness of initial current sheet.
The magnetic Reynolds (Lundquist) number is
\begin{equation}
%
R_{\rm m,y}^{\rm init}
\equiv {v_{\rm A}^{\rm init}l^{\rm init}\over\eta_0}
\sim 167.
\end{equation}

\paragraph{\bf Phase I: Shock Wave Passage ($\it\bf t<2$) and
Current-Sheet Thinning by Tearing Instability ($\it\bf t\sim 100-200$)}

The supernova blast shock (a fast-mode magnetohydrodynamic shock)
crosses the current sheet in an early phase ($t<2.5$; figure [\ref{InitDe}]). 
The magnetic reconnection does not occur 
immediately after the shock wave crosses the current sheet,
because the crossing time is too short to drive the magnetic reconnection.

Figure \ref{Thickness} shows the time variation of
half-thickness of the current sheet,
defined by the minimum half-width of the half-maximum of
gas pressure calculated parallel with $y$-axis in $|x|<25$.
Figure \ref{ThicknessPic} shows its schematic illustration.
The magnetic field is dissipated ($t\sim 0-100$),
and the current-sheet thickness increases slowly.
The current-sheet thinning occurs by the tearing instability
($t\sim 100-200$).
Figure \ref{Flux} shows the time variation of magnetic flux, defined by
$\psi(t)=\int^t_0 \max_{|x|<10}
                 | (B_{\rm x} v_{\rm y})[\rm at ({\it x}, 2)]
                  +(B_{\rm x} v_{\rm y})[\rm at ({\it x},-3)]|$dt. 
The process in this phase corresponds to the 
nonlinear stage of the tearing-mode instability
(\cite{fur63}; \cite{mag99}; Steinolfson and \& Hoven 1983, 1984).
The growth rate is consistent with the results of
Magara \& Shibata (1999).
The dashed line in figure \ref{Flux} shows $\log\psi(t)\sim 0.00257t$.
It is 1/2 times that of the linear phase
[$\log\psi(t)
=    \log_{10}e\times 0.43(1/\tau_{\rm t}^{\rm init})(1/4\pi)t
\sim 0.0042t$; \cite{mag99}].
It explains the result of our numerical simulation well.

The half-thickness of the current sheet decreases to
$l\sim 0.35$ at $t\sim 200$ when the current-sheet thinning stops.
The current-sheet length is
comparable to the most unstable wavelength of the tearing instability
occurring in the initial current sheet, i.e., 
\begin{equation}
\lambda_{\rm t}\sim 5.6R_{\rm m,y}^{\rm init}{}^{1/4}l^{\rm init}
\end{equation}
(\cite{mag99}).
It is $\sim 20$ in this model.
The current-sheet thickness in this phase is explained by
\begin{equation}
l_{\rm t}\sim ({\lambda_{\rm t}\over 2})R_{\rm m,t}^{-1/2},
\label{lct}
\end{equation}
i.e., this current sheet corresponds to Sweet-Parker current sheet (\cite{mag99}).
Here, the magnetic Reynolds number is
\begin{equation}
R_{\rm m,t}={(\lambda_{\rm t}/2)v_{\rm A}^{\rm init}\over\eta_0}.
\label{Rmt}
\end{equation}
The equations (\ref{lct}) and (\ref{Rmt}) give us
$l_{\rm t}\sim 0.23$ and $R_{\rm m,t}\sim 1467$, respectively.
These values explain our results in this phase well.
In summary, we can say that the current-sheet thinning stops
when the current sheet evolves to Sweet-Parker current sheet.
The gas flows into the diffusion region
at the velocity of $v_{\rm in}\sim 0.01$.
The duration of this phase is $\triangle t_{\rm t}\sim 100$,
which is approximately equal to 
the time for initial current sheet to become the Sweet-Parker sheet,
$l^{\rm init}/v_{\rm in}\sim 1/0.01\sim 100\sim 20\tau_{\rm t}^{\rm init}$.

\paragraph{\bf Phase II: Sweet-Parker Type Reconnection
($\it\bf t\sim 200-270$)}

Sweet(1958)-Parker(1957) type reconnection starts
at $t\sim 200$ in an asymmetric situation.
Alfv\'en velocity is higher in the region ($y>0$)
because of the low density gas generated by the supernova.
The gas flows along current sheet, and is accelerated to $v\sim 2.5$,
which is equal to the Alfv\'en velocity near $(x,y)\sim (0,-2)$.
Figure \ref{Inflow} shows
(a) the time variation of velocity of the gas
flowing into the reconnection region.
The inflow velocity is defined by
$v_{\rm in}=\max_{|x|<10}
\left[v_{\rm y}[\rm at ({\it x},-3)]-v_{\rm y}[\rm at ({\it x},1)]\right]/2$.
Figure  \ref{Inflow}(b) shows Alfv\'en velocities attained
at $(x_0,2)$ (the solid line) and at $(x_0,-2)$ (the dashed line)
where $x_0$ is the point where maximum inflow velocity attained.
Figure \ref{Inflow}(c) shows the reconnection rate defined
by $v_{\rm in}/v_{\rm A,s}$,
where $v_{\rm A,s}$ is the smaller one between
$v_{\rm A}[\rm at ({\it x}_0, 1)]$ and $v_{\rm A}[\rm at ({\it x}_0,-3)]$.
The current-sheet length increases to $\lambda\sim 80$
(figure \ref{Thickness}),
because the gas in the initial current sheet flows out
along the current sheet.
The inflow velocity becomes $v_{\rm in}\sim 0.025$.
The ratio of inflow to outflow velocity ($\sim 0.025/2.5\sim0.01$)
is explained by
\begin{equation}
{v_{\rm in}\over v_{\rm A}}
\sim {l_{\rm SP}\over\lambda_{\rm SP}/2}
\sim R_{\rm m,SP}^{-1/2}
\label{SP}
\end{equation}  
where $R_{\rm m,SP}$ is the local magnetic Reynolds number defined by
\begin{equation}
R_{\rm m,SP}={(\lambda_{\rm SP}/2)v_{\rm A}\over\eta_0},
\end{equation}  
$\lambda_{\rm SP}$ and $l_{\rm SP}$ are 
the length and half-thickness of the Sweet-Parker current sheet
(\cite{par57}), respectively,
at the time when the secondary tearing instability starts .
These equations give us $v_{\rm in}/v_{\rm A}\sim 0.01$
with $R_{\rm m,SP}\sim 10^4$, $\lambda_{\rm SP}\sim 80$,
and $l_{\rm SP}\sim 0.3$, respectively, in this model.
The condition for the setup of the secondary tearing instability
is determined by the aspect ratio of Sweet-Parker current sheet,
so that the secondary tearing instability occurs 
when aspect ratio [length/thickness$=(\lambda/2)/l$]
becomes larger than 100 (\cite{bis93}).
The equation (\ref{SP}) explains the result of our simulation 
in this phase ($t\sim 220-250$) well.
%
%
During the Sweet-Parker type reconnection,
the current-sheet thickness increases
in proportion to $\lambda^{1/2}$,
which is derived from the equation (\ref{SP}).

At $t\sim 200-220$,
the inflow velocity ($v_{\rm in}$) increases to $\sim 0.15$
(see the arrow in figure \ref{Inflow}),
which is much larger than 
Sweet-Parker value $\sim R_{\rm SP}^{-1/2}v_{\rm A}\sim 0.025$.
This transient inflow velocity is determined 
by $v_{\rm in}\sim 2l_{210}v_{\rm out}/\lambda_{\rm t}\sim 0.15$,
which is derived from the mass conservation equation
$\lambda_{\rm t}v_{\rm in}\sim 2l_{210}v_{\rm out}$,
assuming the density is constant
and $v_{\rm out}$ ($\sim 0.6$) is the velocity of the outflowing gas
along the current sheet at this time ($t\sim 210$),
where $l_{210}\sim 2.5$ is the current-sheet half-thickness
in $x\sim 20$ (at the edge of Sweet-Parker current sheet)
at $t\sim 210$ (Cf. \cite{fu86}; \cite{shi97}).

Figure \ref{Energycen} shows the time variation of various energies
in the reconnection region ($|x|<25$ and $|y|<3$).
The magnetic energy (the solid line), thermal energy (the dotted line),
and kinetic energy (the dashed line) are defined by
$E_{\rm mag}=\int_{-25}^{25}\int_{-3}^3(B^2/8\pi)$dxdy,
$E_{\rm th}=\int_{-25}^{25}\int_{-3}^3[p_{\rm g}/(\gamma-1)]$dxdy,
and $E_{\rm kin}=\int_{-25}^{25}\int_{-3}^3(\rho v^2/2)$dxdy,
respectively.
They are corrected by the energy flux
at the boundaries ($|x|=25$ and $|y|=3$).
Figure \ref{EnergycenPic} shows the schematic illustration
of the time variation of the magnetic energy-release rate.
The magnetic energy is converted to the thermal energy slowly 
by the tearing instability and Sweet-Parker type reconnection.
At $t\sim 200-220$, the magnetic energy-release rate increases
(see the arrows in figures \ref{Energycen} and \ref{EnergycenPic})
by the nonlinear growth of tearing instability.

\paragraph{\bf Phase III:
Secondary Tearing Instability ($\it\bf t\sim 250-270$)}

The Alfv\'en time and time scale of the tearing instability
of the Sweet-Parker current sheet become,
\begin{eqnarray}
\tau_{\rm A,SP}&=& {l_{\rm SP}\over v_{\rm A}}\sim 0.12,\\
\tau_{\rm t,SP}&=& ({l_{\rm SP}{}^3\over\eta_0 v_{\rm A}})^{1/2}
\sim 0.85,
\end{eqnarray}
respectively.
The tearing instability
(we call ``secondary tearing instability'' in this paper)
occurs in Sweet-Parker current sheet at $t\sim 250$.
The current-sheet thickness decreases again.
Some magnetic islands are generated with intervals of $\sim 5-10$ in $x$,
which are determined by the wavelength of secondary tearing instability,
\begin{equation}
\lambda_2\sim 4.9R_{\rm m,y}^{1/4}l_{\rm SP},
\label{lambda2}
\end{equation}
where $R_{\rm m,y}$ is the local Reynolds number, defined by
\begin{equation}
R_{\rm m,y}={v_{\rm A}l_{\rm SP}\over\eta_0},
\end{equation}
and $v_{\rm A}$ is the local Alfv\'en velocity.
The equation (\ref{lambda2}) gives us $\lambda_2\sim 4$
with $R_{\rm m,y}\sim 50$, which explains the results of our simulation
in this phase well.
Some small islands collide with one another, and are coalesced
to become big islands.
The secondary tearing instability goes on for $\triangle t_{\rm 2t}\sim 20$.
It is equal to $2.5\tau_{\rm t,SP}\sim 20$.

\paragraph{\bf Phase IV:
Petschek Type Reconnection ($\it\bf t\sim 270-400$)}

Figure \ref{Drift} shows the time variations of
(a) maximum drift velocity ($v_{\rm d}\equiv J/\rho$) in the current sheet
 ($|x|<5$ and $-2.6<z<1.0$),
(b) density where maximum $v_{\rm d}$ is attained,
(c) current density ($J$) where maximum $v_{\rm d}$ is attained,
and (d) maximum magnetic reconnection rate defined by $\eta J$.
The density decreases from 0.5 to 0.1 at the center of the current sheet
(figure \ref{Drift}) because the gas flows away along the current sheet
from the dissipation region.
The maximum drift velocity exceeds the threshold ($v_{\rm c}=100$)
in the reconnection region
immediately after the big magnetic island moves away.
The anomalous resistivity sets in at $t\sim 270$.
The half-thickness of the current sheet decreases to $l\sim 0.15$
(figure \ref{Thickness}).
The diffusion region is much localized near the X-point.
Petschek (1964) type reconnection begins
with various violent phenomena.
The gas flows into reconnection region
at $v_{\rm in}\sim 0.1-0.2$ (figure \ref{Inflow}).
The magnetic reconnection rate (figure \ref{Inflow}) increases to
\begin{equation}
\epsilon={v_{\rm in}\over v_{\rm A}}\sim 0.05-0.10,
\label{epsilon}
\end{equation}
while it is $\sim 0.02$
during Sweet-Parker type reconnection.
In this simulation,
the energy-release rate during Petschek type reconnection is
only a few times higher than that during Sweet-Parker type reconnection.
The latter, however, is much small in the actual Galaxy
because it is determined by the magnetic Reynolds number
as $\epsilon=v_{\rm in}/v_{\rm a}=R_{\rm m}^{-1/2}$.
The gas in the diffusion region are accelerated along the current sheet
by magnetic tension force to $v\sim 2.5-3$,
which is comparable to the Alfv\'en velocity in the region
where the maximum velocity of reconnection jet is attained.

The magnetic energy is converted to the thermal energy slowly 
during the tearing instability and Sweet-Parker type reconnection,
and quickly during Petschek type reconnection,
particularly while magnetic islands are moving 
along the current sheet from the reconnection region
(figure \ref{Energycen}).
The energy conversion is suppressed when 
the magnetic islands are staying in the diffusion region
($t\sim 310$ and 360).
Total released magnetic energy is $\sim 5500$ 
in the reconnection region ($|x|<25$ and $|y|<3$; figure \ref{Energycen}),
which is much higher than the initial thermal energy of the supernova
($E_{\rm ex}\sim 1440$).
The maximum energy release rate and maximum heating rate are
$-dE_{\rm mag}/dt\sim dE_{\rm th} /dt\sim 70-80$ at $t\sim 340$.
They are determined by the Poynting flux entering the reconnection region,
\begin{equation}
-{dE_{\rm mag}\over dt}
\sim 2{B^2\over 4\pi}Sv_{\rm in},
\end{equation}
where $S$ is the reconnection region size.
The above equation gives us $\sim 80$,
which explains the results of the simulation
(The reconnection region size is derived as $S\sim 37$
if we use the values of $B\sim 8$ and $v_{\rm in}\sim 0.2$.).

The current density increases in the current sheet
(at $t=210$ and 250; figure \ref{Current})
and behind the moving magnetic islands 
(at $t=275$ and 300; figure \ref{Current}).
The slow shocks are formed behind the moving islands.
Two standing slow shocks are generated in the current sheet
(at $t=325$ and 350; figure \ref{Current}).
Figure \ref{SShock} shows the profile of the various physical quantities
in $x=-12$ at $t=325$.
The gas pressure, temperature, density, and velocity
are high in the current sheet.
The magnetic field is weak in the current sheet.
The current density is large at the slow shock regions
so that the profile of current density has double peaks.
It is noted that this Petschek-type reconnection
is very time-dependent;
Some magnetic islands are repeatedly generated
and ejected in the current sheet,
and magnetic energy-release rate is controlled by the magnetic islands.

\paragraph{\bf Phase V:
End of Fast (Petschek Type) Reconnection ($\it\bf t\sim 400$)}

The current-sheet length is limited by the high gas pressure
near the right and left symmetric boundaries ($x=-120.8$ and $x=120.8$).
Figure \ref{Whole} shows the distribution of
gas pressure in whole simulation region.
The current sheet, whose half-thickness was $l^{\rm init}=1$ at $t=0$,
expands near the left and right boundaries.
The magnetic loops are accumulated at the left and right boundaries
so that current-sheet thickness increases
from $\sim 2$ at $t=160$ (figure \ref{Whole}[a])
to $\sim 20$ at $t=300$ (figure \ref{Whole}[b]),
and eventually to $\sim 50$ at $t=450$ (figure \ref{Whole}[c]).
The gas pressure increases in the reconnected loop.
The current-sheet length decreases to $\lambda\sim 50$ at $t\sim 450$
(figure \ref{Whole}[c]).
The ejection from the diffusion region is suppressed
by the high gas pressure in the loop near the boundaries,
so that the global field configuration tends to be the potential field
produced by the localized currents near the boundaries.
On the other hand, the magnetic tension force suppresses
the inflow to reconnection region
because the potential magnetic field outside the current sheet is bent.
Consequently, the magnetic reconnection rate decreases suddenly
at $t \sim 380$.
Duration of the Petschek type reconnection is
$\triangle t_{\rm P}\sim 170$.

\subsection{Parameter Survey}

\subsubsection{Dependence of Results on Magnetic Field Strength ($\beta$)}

We examine the dependence of results on the magnetic field strength
[$B_0=(8\pi p_0/\beta)^{1/2}$], i.e., $\beta$ (Model A2 and B1-5).
The time variations of magnetic energy-release rates
are shown in figure \ref{Emag_beta}
by the dotted line (Model B1; $\beta=0.1$),
solid line (Model A2; $\beta=0.2$),
and dashed line (Model B2; $\beta=0.3$).
The magnetic energy-release rate decreases with $\beta$
as $\sim 250$ ($\beta=0.1$), $\sim 70$ ($\beta=0.2$),
and $\sim 20$ ($\beta=0.3$), respectively.

Figure \ref{Beta} (a) shows the $\beta$-dependence of
maximum magnetic energy release rates
during Petschek type reconnection (the solid line)
and during Sweet-Parker type reconnection (the dotted-line)
in the reconnection region ($|x|<25$ and $|y|<3$).
The energy-release rate decreases with $\beta$.
The rate during Petschek type reconnection is determined by
the Poynting flux entering into reconnection region,
\begin{equation}
-{dE_{\rm mag}\over dt}|_{\rm P}
\sim   2{B^2\over 4\pi}Sv_{\rm in}
\propto\beta^{-3/2},
\label{dedt}
\end{equation}
where $S$ is the reconnection region size,
$v_{\rm in}=\epsilon v_{\rm A}=\epsilon B/(4\pi\rho)^{1/2}$
is the inflow velocity to the reconnection region,
and $\epsilon$ is the reconnection rate.
This theoretical $\beta$-dependence (\ref{dedt})
is also shown in the figure \ref{Beta}(a) by the dashed line
[$\sim 2(B_0^2/4\pi)S\epsilon v_{\rm A}^{\rm init}
\sim 6.8\beta^{-3/2}$ where $S$ is derived to be $\sim 26$
if we use the results of the typical model].
It explains the numerical results well.
On the other hand, 
the magnetic energy-release rate during Sweet-Parker reconnection
is given by
\begin{equation}
-{dE_{\rm mag}\over dt}|_{\rm SP}
\sim   2{B^2\over 4\pi}Sv_{\rm in}
\propto\beta^{-5/4},
\label{dedtSP}
\end{equation}
where $S$ is the reconnection region size,
$v_{\rm in}=R_{\rm m,SP}^{-1/2}v_{\rm A}\propto v_{\rm A}^{1/2}$
is the inflow velocity to the reconnection region.
This theoretical $\beta$-dependence (\ref{dedtSP})
is also shown in the figure \ref{Beta}(a) by the dashed-dotted line
[$\sim 2(B_0^2/4\pi)SR_{\rm m,SP}^{-1/2}v_{\rm A}^{\rm init}
\sim 2.5\beta^{-5/4}$ where $S$ is derived to be $\sim 67$
if we use the results of the typical model].
It explains the numerical results well.

Figure \ref{Beta}(b) presents the times
when maximum magnetic energy release rates are attained
during Petschek (the solid line)
and Sweet-Parker (the dashed line) type reconnections.
The time when Sweet-Parker type reconnection starts
is determined by the time scale of dissipation in an early phase
($\sim\triangle t_{\rm t}$)
plus the time scale of tearing instability ($\sim\triangle t_{\rm t}$).
It is almost equal to $\sim 2\triangle t_{\rm t}$.
The time scale of tearing instability
($\tau_{\rm t}\propto v_{\rm A}^{-1/2}\propto\beta^{1/4}$)
is presented in figure \ref{Beta}(b) by
the dashed-dotted line ($\sim 230\beta^{1/4}$).
The time when Petschek type reconnection starts
is determined by the time when Sweet-Parker type reconnection
starts ($\sim 2\triangle t_{\rm t}$) plus
the time for the Sweet-Parker current sheet to get long
until it becomes unstable to the secondary tearing instability,
$\lambda/2\geq 100l_{\rm SP}\sim 30$ (\cite{bis93}),
which is equal to $\sim \lambda_{\rm SP}/v_{\rm A}$.
The Alfv\'en time
($\tau_{\rm A}\propto v_{\rm A}^{-1}\propto\beta^{1/2}$)
is presented in figure \ref{Beta}(b) by
the dashed line ($\sim 580\beta^{1/2}$).

Figures \ref{Beta}(c) and (d) show
the maximum temperature of heated gas
and maximum velocity of the reconnection jet ($|y|\leq 2$).
The maximum temperature and the maximum velocity decreases with $\beta$.
The interstellar gas is heated, by passing the reconnection region, to
\begin{equation}
T_{\rm max}
\sim\left(1+{1\over\beta}\right)
    \left({n_{\rm in}\over n_{\rm out}}\right)T_{\rm in}
\propto\left(1+{1\over\beta}\right),
\label{tmax}
\end{equation}
which is derived from the equation
$n_{\rm out}kT_{\rm max}\sim n_{\rm in}kT_{\rm in}+B^2/8\pi$.
Here, we assume $(n_{\rm in}/n_{\rm out})T_{\rm in}$
is independent of $\beta$.
The $\beta$-dependence is plotted in figure \ref{Beta}(c)
by the dashed line [$\sim 7(1+1/\beta)$].
In the simulation,
a part of hot gas of the supernova flows into the reconnection region,
and re-heated by the reconnection.
The gas is accelerated to Alfv\'en velocity,
\begin{equation}
v_{\rm jet}\sim
v_{\rm A}={B\over (4\pi\rho)^{1/2}}\propto\beta^{-1/2},
\end {equation}
by the magnetic tension force along the current sheet,
which is shown in the figure \ref{Beta}(d) by the dashed line
($\sim 1.5\beta^{-1/2}$
which is 1.5 times higher than $v_{\rm A}^{\rm init}$
because $v_{\rm A}$ increases with time).

\subsubsection{Dependence of Results on Distance
Between Supernova and Current Sheet}

We examine the dependence of results on the distance ($y_{\rm ex}$)
between the supernova and current sheet (Models A2 and C1-3).
Figures \ref{Distance}(a) and \ref{Distance}(b) illustrate
the maximum magnetic energy release rate
and the time when it is attained, respectively.
The magnetic energy-release rate decreases with $y_{\rm ex}$.
The time scale increases with $y_{\rm ex}$.
It is because the initial perturbation decreases with $y_{\rm ex}$,
the time when magnetic reconnection starts increases
and more magnetic field dissipates before the reconnection starts.
The maximum temperature (figure \ref{Distance}[c])
decreases with $y_{\rm ex}$.
The maximum velocity of reconnection jet (figure \ref{Distance}[d])
does not depend on the distance
because it is determined by the Alfv\'en velocity.

\subsubsection{Dependence of Results on Initial Energy of Explosion}

We put the point explosion with the gas pressure of
$E_{\rm ex}\propto p_{\rm ex}=p_{\rm ratio}p_0$ as the initial condition.
We examine the dependence of results on $p_{\rm ratio}\propto E_{\rm ex}$
(Models A2 and D1-5; figure \ref{Pratio}).
The normalizations given by the equations (\ref{H})-(\ref{tau})
change if $p_{\rm ratio}$ of the equation (\ref{Esn}) changes.
We must rewrite the units of length and time as
$H\propto p_{\rm ratio}^{-1/3}$
and $\tau\propto p_{\rm ratio}^{-1/3}$, respectively.
Figure \ref{Pratio}(a) presents the maximum magnetic energy release rate.
It does not depend on the energy of initial explosion.
Figure \ref{Pratio}(b) shows the time
when maximum magnetic energy release rate is attained.
It decreases with $p_{\rm ratio}$
because the initial perturbation decreases.
Figure \ref{Pratio}(c) shows the maximum temperature of gas
heated by the reconnection.
The temperature increases with $p_{\rm ratio}$
because the hot gas of the initial supernova is re-heated
by the magnetic reconnection.
Figure \ref{Pratio}(d) shows the velocity of  reconnection jet.
It does not depend on $p_{\rm ratio}$
because it is determined by the Alfv\'en velocity.

We examine the model
in which two explosions appear symmetrically
in both sides [at $(0,7)$ and $(0,-7)$] of the current sheet
($p_{\rm ratio}=500$; Model D5).
The results are plotted in figure \ref{Beta} by the cross ($\times$).
The time scale is much shorter
because the initial current sheet is compressed from both sides.
Other results are similar to those of typical model.

\subsubsection{Dependence of Results on Resistivity Model}

We examine the dependence of the results
on the background uniform resistivity ($\eta_0$) 
and the resistivity models 
(either anomalous or uniform for whole evolutions).
Figure \ref{Thickness_eta} shows
the time variations of the half-thickness of the current sheet,
for Model A2 ($\eta_0=0.015$; thick solid line),
Model F1 ($\eta_0=0.03$; thin dotted line),
Model F2 ($\eta_0=0.0075$; thin dashed-dotted line),
Model F3 ($\eta_0=0.00375$; thin dashed-dotted-dotted line),
Model F4 ($\eta_0=0.001$; thin dashed line),
Model F5 ($\eta_0=0.0001$; thin solid line),
and Model F6 ($\eta=0$; thick dashed line).
In the current sheet of Model A2 ($\eta_0=0.015$),
the tearing instability ($t\sim 60-150$) is followed by
Sweet-Parker type reconnection ($t\sim 150-200$),
secondary tearing instability ($t\sim 200-240$),
and Petschek type reconnection ($t\sim 240$).
In zero-resistivity model (Model F6),
the current sheet does not dissipate at first.
It, however, starts to become thin at $t\sim 220$,
and eventually leads to Petschek type reconnection
by numerical resistivity (resistivity-like effect by numerical noise).
The result of Model F5 ($\eta_0=0.0001$) is similar to
that of Model F6 ($\eta=0$), not that of F4 ($\eta_0=0.001$).
The numerical resistivity in these simulations is estimated
as $\eta_{\rm num}=0.001-0.0001$.
In Model F1 ($\eta=0.03$), the current sheet dissipates ($t<100$),
and the magnetic reconnection does not occur ($t<800$).
The maximum temperature of the gas heated by the reconnection
and the velocity of the reconnection jet do not depend on $\eta_0$.

The result of uniform resistivity model (Model E)
is also shown in Figure\ref{Beta} by triangle ($\triangle$).
Petschek type reconnection does not occur
in the uniform resistivity model.
The magnetic energy-release rate and the time scale 
are approximately equal to the results of anomalous resistivity model
during the Sweet-Parker type reconnection.

\subsubsection{Dependence of Results of Simulation Region Size and Grid Size}

We examine the dependence of results on the simulation region size ($L_{\rm y}$)
in $y$ axis (Models A2 and G1-2).
The magnetic energy release rate increases with the simulation region size
because the stored initial total magnetic energy increases.
The other results do not depend on the simulation region size.

We examine the effects of grid size ($\triangle x$) (Models A1-2). 
Until the secondary tearing instability starts,
the time evolution of current sheet depends on the parameters
such as $\beta$ and $\eta$ well
so that the time scales of phenomena are predictable.
At this stage (the onset of the secondary tearing instability), however, 
it becomes difficult to predict where small islands appear
and when and how they move and collide with one another,
since the current-sheet thickness decreases to a few $\times\triangle y$
after the secondary tearing instability starts.
At this stage, the current sheet is too thin to be free from 
numerical resistivity in our problem.
The maximum energy release rate, maximum temperature of the heated gas
and velocity of the reconnection jet do not depend on $\triangle x$,
although the time when maximum magnetic energy release rate is attained
depends on $\triangle x$ (figure \ref{Emag_grid}).

\section{DISCUSSION}

Odstr\v{c}il \& Karlick\'{y} (1997) performed an interesting
2D numerical simulation of the magnetic reconnection triggered by
a point explosion near the current sheet,
with parameters different form ours,
in order to construct a model of the solar flare triggered by a distant flare.
They assumed resistivity as $\eta\propto |J-J_{\rm c}|$,
where $J_{\rm c}$ is the critical current density
above which the anomalous resistivity sets in.
Zimmer et al.\ (1997) tried to explain
that the gas is heated in Galactic halo by magnetic reconnection
triggered by the collision of the high-velocity clouds
(HVC; $M\sim 10^{7-8}M_\odot$; $V\sim 100$ km s$^{-1}$; \cite{bli99}),
by performing the 2D MHD numerical simulation.
Birk et al.\ (1998) performed 2D numerical simulation of
the magnetic reconnection including the ionization and recombination.
Above three groups
followed the current-sheet thinning and the generation of the magnetic island.
They, however, did not reveal the time evolution of the secondary tearing instability,
because they did not use enough small grid size.
In this paper, we use 80 grid points,
so that we revealed how the secondary tearing instability leads
to Petschek type magnetic reconnection.

We revealed how the fast reconnection starts at the current sheet
via the secondary tearing instability
by assuming the anomalous resistivity model (equation [\ref{eta}]) in this paper.
Main result of this paper is revealing how the current sheet
leads to fast reconnection via the secondary tearing instability.
This result does not depend on the form of resistivity before
the anomalous resistivity sets in (\cite{yok95}; \cite{uga92}).
The simulation results we apply to the actual Galaxy,
such as temperature of heated gas, do not depend on 
the resistivity model (\cite{yok95}; \cite{uga92}).

We assume that the initial current-sheet has enhanced gas pressure
[$p_{\rm g}=p_0+(p_0/\beta)\cosh^{-2}(y)$;
$B_{\rm x}=B_0\tanh(y)$; $B_{\rm y}=0$] in the paper.
The tearing instability grows similarly
in both this model (enhanced gas pressure model) and force-free model
[$p_{\rm g}=p_0$;
$B_{\rm x}=B_0\cos 0.5\pi(y+1)$;
$B_{\rm z}=B_0\sin 0.5\pi(y+1)$; $B_{\rm y}=0$; \cite{mag99}].
The basic physics of magnetic reconnection do not change,
though the nonlinear phase starts earlier
in the enhanced gas pressure models (\cite{mag99}).
The physical processes of our results are hardly influenced
by the boundary conditions
because the simulation region is large enough.
We do not assume the initially enhanced resistivity in the current sheet
to see how the spontaneous reconnection occurs
after the secondary tearing instability.
We use so small grid size that we can suppress the numerical resistivity
in our numerical simulations.

The results do not depend on $p_{\rm ratio}$ very much.
The magnetic reconnection occurs with any $p_{\rm ratio}$.
It means that the magnetic reconnection occurs
not only by a supernova, but also by the small perturbation.
The magnetic reconnection can be triggered by various mechanisms
such as a supernova (or a point explosion), 
superbubbles (sum of supernovae; \cite{tom98}; \cite{ten88}),
stellar winds, collision of interstellar clouds,
shock waves, Parker(1966) instability
(undular mode of magnetic buoyancy instability),
magnetorotational instability (\cite{bal91}),
collision of galaxies, cosmic rays etc. in numerous current sheets.

The random field stores the magnetic energy at $\sim$ several 10 pc
(\cite{ran89}).
Total magnetic energy is
$E_{\rm mag}
\sim \left[\langle B\rangle_{\rm obs}^2/8\pi\right] V_{\rm G}
\sim 10^{54.4}$  erg,
where $\langle B\rangle_{\rm obs}\sim 3\ \mu$G
is assumed as the mean observed field strength,
and $V_{\rm G}\sim 2\times 10^2$ kpc$^3$
is the volume of Galaxy.
The thermal energies of GRXE (Galactic Ridge X-ray Emission)
and coronal halo are
$E_{\rm GRXE}\sim 10^{55}$ erg (e.g. \cite{kan97})
and $E_{\rm halo}\sim 10^{57.2}$ erg
(\cite{hab80}; \cite{li92}; \cite{ike88}),
respectively.
Total rotational energy of Galaxy ($E_{\rm rot}\sim 10^{58.9}$ erg)
and total kinetic energy of the interstellar gas
($E_{\rm ISM}\sim 10^{58.2}$ erg) 
are higher than the thermal and magnetic energies (\cite{stu80}).
The rotational energy can be converted to the thermal energy
by Galactic dynamo and magnetic reconnection
in $\sim 10^8$ yr (e.g. \cite{par71}; \cite{tan99}).

\subsection{Implication for Observations}

In our numerical simulations,
the magnetic Reynolds number is much smaller than
actual value in Galaxy. 
The Alfv\'en velocity, resistivity, and magnetic Reynolds number are
$v_{\rm A}\sim 6\times 10^6
(B/10\ \mu {\rm G})(n/0.1\ {\rm cm}^{-3})^{-1/2}$ cm s$^{-1}$,
$\eta\sim 10^7(T/10^4\ {\rm K})^{-3/2}$ cm$^2$ s$^{-1}$,
and $R_{\rm m}\sim \lambda v_{\rm A}/\eta
\sim 2\times 10^{20}(B/10\ \mu{\rm G})(n/0.1\ {\rm cm}^{-3})^{-1/2}
(\lambda/100\ {\rm pc})(T/10^4\ {\rm K})^{3/2}$,
if we assume the length of current sheet
which is getting thin by the tearing instability,
where the local magnetic field strength,
number density, and temperature are
$\lambda\sim 100$ pc, $B\sim 10\ \mu$G, $n\sim 0.1$ cm$^{-3}$,
and $T\sim 10^4$ K, respectively.
Sweet-Parker current-sheet thickness is 
\begin{eqnarray}
l_{\rm SP}
&\sim&\lambda R_{\rm m}^{-1/2}\nonumber\\
&\sim& 2\times 10^{10}({B\over 10\ \mu {\rm G}})^{-1/2}
({n\over 0.1\ {\rm cm}^{-3}})^{1/4}\nonumber\\
& &
({\lambda\over 100\ {\rm pc}})^{1/2}
({T\over 10^4\ {\rm K}})^{-3/4}\ \rm cm.
\end{eqnarray}
The time scale of Sweet-Parker reconnection is
\begin{eqnarray}
\tau_{\rm SP}\sim {\lambda\over v_{\rm in}}
&\sim& {\lambda\over v_{\rm A}R_{\rm m}^{-1/2}}\nonumber\\
&\sim& 3\times 10^{16}({\lambda\over 100\ {\rm pc}})^{3/2}
({B\over 10\ \mu{\rm G}})^{-1/2}\nonumber\\
& &({n\over 0.1\ {\rm cm}^{-3}})^{-3/4}
({T\over 10^4\ {\rm K}})^{3/4}\ \rm yr
\end{eqnarray}
(under classical Spitzer resistivity),
so that the Sweet-Parker type reconnection can not play an important role
as energy-release mechanism in actual interstellar medium.

On the other hand, the time scales of the secondary tearing instability
and current-sheet thinning are
\begin{eqnarray}
\tau_{\rm 2,tear}
&\sim& ({l^3\over\eta v_{\rm A}})^{1/2}
\sim ({l\over\lambda})^{3/2}({\lambda\over v_{\rm A}})R_{\rm m}^{1/2},\\
\tau_{\rm thin}
&\sim& {\lambda\over v_{\rm A}}\nonumber\\
&\sim& 2\times 10^6({\lambda\over 100\ {\rm pc}})\nonumber\\
& &({B\over 10\ \mu{\rm G}})^{-1}
({n\over 0.1\ {\rm cm}^{-3}})^{1/2}\ \rm yr,
\end{eqnarray}
respectively.
The secondary tearing instability occurs faster than
the current-sheet thinning
if $\tau_{\rm 2,tear}<\tau_{\rm thin}$, i.e.,
the current-sheet thickness decreases to
\begin{eqnarray}
l &<&  l_{\rm 2,tear}\nonumber\\
  &=&  \lambda R_{\rm m}^{-1/3}\nonumber\\
&\sim& 5\times 10^{13}({B\over 10\ \mu {\rm G}})^{-1/3}
({n\over 0.1\ {\rm cm}^{-3}})^{1/6}\nonumber\\
& &({\lambda\over 100\ {\rm pc}})^{2/3}
({T\over 10^4\ {\rm K}})^{-1/2}\ \rm cm.
\end{eqnarray}
In this situation,
the aspect ratio becomes $\lambda/(2l_{\rm tear})\sim 4\times 10^6$.
Hence, the secondary tearing instability starts
before Sweet-Parker current sheet develops completely.
The time scale and wavelength of the secondary tearing instability are
calculated to be
\begin{eqnarray}
\tau_{\rm 2,tear}
&\sim& 2\times 10^6({B\over 10\ \mu {\rm G}})^{-1/3}
({n\over 0.1\ {\rm cm}^{-3}})^{1/6}\nonumber\\
& & ({\lambda\over 100\ {\rm pc}})^{2/3}
({T\over 10^4\ {\rm K}})^{-1/2}\ \rm yr.\\
\lambda_{\rm 2,tear}
&\sim& 6({v_{\rm A}l_{\rm 2,tear}\over\eta})^{1/4}l_{\rm 2,tear}
\nonumber\\
&\sim& 6\lambda R_{\rm m}^{-1/6}\nonumber\\
&\sim& 1\times 10^{18}({B\over 10\ \mu {\rm G}})^{-1/6}
({n\over 0.1\ {\rm cm}^{-3}})^{1/12}\nonumber\\
& & ({\lambda\over 100\ {\rm pc}})^{5/6}
({T\over 10^4\ {\rm K}})^{-1/4}\ \rm cm,
\end{eqnarray}
respectively.

When a plasmoid moves along the current sheet,
the inflow velocity increases behind it so that
the current-sheet width decreases to
\begin{eqnarray}
l_{\rm min}
&\sim& {\eta\over v_{\rm in}}\nonumber\\
&\sim&\lambda R_{\rm m}^{-5/6}\nonumber\\
&\sim& 3\times 10^3({B\over 10\ \mu{\rm G}})^{-5/6}
({n\over 0.1\ {\rm cm}^{-3}})^{5/12}\nonumber\\
& & ({\lambda\over 100\ {\rm pc}})^{1/6}
({T\over 10^4\ {\rm K}})^{15/12}\ \rm cm,
\end{eqnarray}
by assuming $\lambda_{\rm 2,tear}v_{\rm in}=l_{\rm 2,tear}v_{\rm A}$,
where $v_{\rm in}$ is the inflow velocity toward the dissipation region.
%
In this transient large inflow stage,
the anomalous resistivity can set in,
which eventually leads to Petschek type reconnection,
since the condition of anomalous resistivity is that
the current-sheet thickness must be smaller than the ion Lamor radius
[$\sim 10^7(T/10^4\ {\rm K})^{1/2}(B/10\ \mu {\rm G})^{-1}$ cm;
figure \ref{Actual}].
The current-sheet thickness, otherwise, decreases to the ion Lamor radius
through third, forth, and fifth tearing instabilities (figure \ref{Actual}).
The third tearing instability occurs
when the current-sheet thickness becomes $\sim 1\times 10^{12}$ cm
(wavelength is $\sim 7\times 10^{15}$ cm;
the time scale of the third tearing instability is $\sim 10^4$ yr)
in the current sheet generated by the secondary tearing instability.
The forth, fifth, sixth, seventh, and eighth tearing instabilities start 
when the current-sheet thickness becomes
$\sim 4\times 10^{11}$ cm, $\sim 2\times 10^{10}$ cm, $\sim 1\times 10^9$ cm,
$\sim 7\times 10^7$ cm, and $\sim 4\times 10^6$ cm, respectively.
Total released magnetic energy is
\begin{equation}
E\sim ({B^2\over 8\pi})\lambda^3
\sim 10^{50}({B\over 10\ \mu{\rm G}})^2
({\lambda\over 100\ {\rm pc}})^3\ \rm erg,
\end{equation}
if we assume the volume of space filled with the magnetic field is $\lambda^3$.
Petschek reconnection continues for
\begin{eqnarray}
\tau_{\rm P}
&\sim& {\lambda\over\epsilon v_{\rm A}}\nonumber\\
&\sim& 10^7({\epsilon\over 0.1})^{-1}({\lambda\over 100\ {\rm pc}})
\nonumber\\
& & ({B\over 10\ \mu{\rm G}})^{-1}
({n\over 0.1\ {\rm cm}^{-3}})^{1/2}\ \rm yr,
\end{eqnarray}
which is equal to one-third of the time of Galactic rotation.
%
The magnetic energy-release rate is 
\begin{eqnarray}
{E\over\tau_{\rm P}}&\sim& 10^{34.5}({\epsilon\over 0.1})^{-1}
({\lambda\over 100\ {\rm pc}})^2\nonumber\\
& &({B\over 10\ \mu{\rm G}})^3
({n\over 0.1\ {\rm cm}^{-3}})^{-1/2}\ \rm erg\ s^{-1}.
\end{eqnarray}

\subsection{Generation of X-ray Gas in Galaxy\label{grxe}}

Figure \ref{Xray} shows the 2D distribution of
the thermal radiation [$F\propto n^2\Lambda(T)$],
for the typical model (Model A1),
where $\Lambda(T)$ is the cooling function ($\sim T^a$).
The above $a$ is $0.5$, $-0.6$, $2.9$, and $1.5$, respectively
when the unit of temperature is $T_0\sim 10^7$ K (figure \ref{Xray}[a]),
$\sim 10^{5-6}$ K (figure \ref{Xray}[b]),
$10^4$ K (figure \ref{Xray}[c]), and $10^3$ K (figure \ref{Xray}[d]).
The X-ray thermal emission shown
in figures \ref{Xray}(a) and \ref{Xray}(b)
may be observed by {\it Newton} and {\it Chandra} in the galaxies.

The bright thermal X-ray emission are observed along the Galactic plane
($L_{\rm X}\sim 10^{38}$ erg s$^{-1}$; Galactic Ridge X-ray Emission=GRXE;
e.g. \cite{kan97}; \cite{koy86a}; \cite{mak94};
\cite{yam97}; \cite{yam96}; \cite{val98}).
If the magnetic reconnection occurs in the magnetic field of $B\sim 30\ \mu$G,
which is generated locally by Galactic dynamo,
the hot ($T\sim 10^8$ K) component ($n\sim 3\times 10^{-3}$ cm$^{-3}$;
\cite{kan97}) of GRXE can be generated (\cite{mak94}; \cite{tan99}).
After then, the hot gas is confined by the locally strong magnetic field for
$\tau_{\rm cond}\sim 10^5(\lambda_{\rm eff}/1\ {\rm kpc})
(T/10^8\ {\rm K})^{-1/2}$ yr, where $\lambda_{\rm eff}$ is
the effective length of helical magnetic field.
Note that the radiative cooling time is
$\tau_{\rm rad}\sim
10^{10.5}(T/10^8\ {\rm K})(n/0.003\ {\rm cm}^{-3})^{-1}$
$[\Lambda(10^8\ {\rm K})/10^{-23}\ {\rm erg\ cm^3\ s^{-1}}]^{-1}$ yr,
which is too much longer than $\tau_{\rm cond}$
so that the radiation is negligible,
where $\Lambda(T)$ is the cooling function.
The heating time by Petschek type reconnection is
$\tau_{\rm P}\sim
5\times 10^5(\lambda/100\ {\rm pc})(B/30\ \mu{\rm G})^{-1}
(n/0.003\ {\rm cm^{-3}})^{1/2}$ yr,
so that the heating is balanced with the conduction cooling,
where $\lambda$ is the thickness of the reconnecting magnetic field.

The hot gas are also observed in Galactic halo (\cite{pie98}),
Galactic center (e.g. \cite{koy89}),
and clusters of galaxies (e.g. \cite{sar86}).
The magnetic reconnection can generate the hot gas
(\cite{zim97}; \cite{bir98}; \cite{ker94}; \cite{ker96}; \cite{sok97})
and confine it in the magnetic loop for a long time.

The reconnection jet is accelerated in opposite directions
by the magnetic tension force, which is like bipolar-jet.
Inside the helical tube and current sheet,
the temperature and gas pressure 
increases by the self-pinch of helical magnetic tube
ot the collision between the reconnection jet and interstellar cloud
so that the star formation may be triggered.

\section{SUMMARY}

We examined the magnetic reconnection triggered by a supernova
(or a point explosion) in interstellar medium,
by performing two-dimensional resistive magnetohydrodynamic (MHD)
numerical simulations with high resolution 
and large simulation region size.
We found that the magnetic reconnection starts long
after a supernova shock (fast-mode MHD shock) passes a current sheet.
The long current sheet evolves as follows:
(i)
The supernova shock perturbs the current sheet.
The magnetic reconnection does not occur immediately after 
the passage of the supernova shock across the current sheet,
because the magnetic field lines do not have enough time to reconnect.
The tearing-mode instability is excited by the supernova shock,
and the current sheet becomes thin in its nonlinear stage.
(ii) The current-sheet thinning is saturated
when the current-sheet thickness becomes comparable to that of Sweet-Parker current sheet.
After that, Sweet-Parker type reconnection starts.
(iii) The current-sheet length increases
during the Sweet-Parker type reconnection,
because the gas in the Sweet-Parker current sheet
flows out along the current sheet.
The secondary tearing mode instability occurs
in the Sweet-Parker current sheet,
because the ratio of length to thickness of the current sheet
exceeds $\sim 100$.
(iv) Further current-sheet thinning occurs
after the islands moved away.
The anomalous resistivity sets in,
and Petschek type reconnection starts.
The magnetic energy is released while the magnetic islands are moving
during the Petschek type reconnection.
The phenomena depend only on $\beta$ (i.e. $B$),
not on the distance between the supernova and current sheet,
nor the energy of initial explosion.

In actual Galaxy, the magnetic reconnection can occur
when the non-parallel magnetic fields (i.e. current sheet) exist
and a point explosion such as supernova, superbubble,
collision of the clouds or stellar wind
will occur near the current sheet.
The initial current sheet will be made by
the turbulence, Parker(1966) instability (\cite{tan98}),
magnetorotational instability (\cite{bal91}),
collision of galaxies etc.
The magnetic islands (or helical tubes) confine the hot gas,
and flow fast along the current sheet.
The interstellar gas flows also inside the magnetic helical tube
in three-dimensional (3D) view.
The gas pressure is very high inside the magnetic islands
so that the gas flows along the tube actually.
In the magnetic tube and current sheet,
the temperature and gas pressure are very high
so that the star forming region may appear.
We will apply the results to
the X-rays from the Galactic ridge, Galactic halo,
Galactic center and clusters of galaxies.
We will examine the reconnection in 3D simulations in future works.

\acknowledgments

The authors thank K. Makishima in University of Tokyo,
R. Matsumoto in University of Chiba,
and T. Magara in Kyoto University for various fruitful discussion.
The numerical computations were carried out on VPP300/16R and VX/4R
at the National Astronomical Observatory of Japan.

\begin{figure}

\vspace{1mm}
\begin{center}
\caption{The situation of our problem.
The interstellar space is filled with the magnetic fields.
The magnetic fields are not exactly parallel to one another,
and have numerous current sheets at all scales.
We examine how the magnetic reconnection occurs
when a supernova perturbs the current sheet.
We calculate the interaction between the supernova and current sheet
by assuming simulation region shown by black square.
\label{Situation}}
\end{center}

\begin{center}
\caption{
Initial condition of the typical model (Model A1) of
our numerical simulations.
We assume the uniform temperature
and uniform total pressure ($P_{\rm g}+B^2/8\pi$) everywhere.
The current-sheet thickness is $2l^{\rm init}H=2H$
where $H$ is the unif of length.
The units of velocity and time are
sound velocity ($C_{\rm S}$) and $H/C_{\rm S}$, respectively.
We solve the non-dimensional equations numerically in this paper.
The magnetic field is assumed as $B_{\rm x}=B_0\tanh(y/H)$ and $B_{\rm y}=0$.
The plasma $\beta$ is $8\pi p_0/B_0^2=0.2$ everywhere
outside the current sheet.
The distance between the supernova and current sheet is $y_{\rm ex}H=7H$.
Minimum grid size is $(\triangle x, \triangle y)\ge (0.20H, 0.025H)$.
The number of grid points is $(465\times 602)$.
The simulation region is $-120.9H\le x\le 120.9H$ and $-54.8H\le y\le 81.7H$.
We assume no gravity ($g=0$).
\label{Init}}
\end{center}

\begin{center}
\caption{
Time evolution of 2D distribution of the current density ($J$),
density ($\rho$), and gas pressure ($p_{\rm g}$), with the magnetic field lines,
in an early phase for the typical model (Model A1).
The half-thickness of initial current sheet is $l^{\rm init}=1$ ($|y|<1$).
The initial density, gas pressure, and magnetic field outside the current sheet
are $n_0=1$, $p_0=0.6$, and $B_0\sim 8.7$, respectively.
The plasma $\beta$ is $8\pi p_0/B_0^2=0.2$ ($|y|>1$).
The Alfv\'en velocity is $B_0/(4\pi\rho)^{-1/2}\sim 2.5$.
We put a supernova at $(0,y_{\rm ex})=(0,7)$.
The supernova expands in the magnetic field.
These figures show the expanding supernova shock well.
The slow shock propagates along the magnetic field.
The shock front is in a high-density region,
while the center of supernova is in a low-density region.
\label{InitDe}}
\end{center}

\end{figure}
\begin{figure}

\begin{center}
\vspace{1mm}
\caption{
Time evolution of 2D distribution of  
the temperature ($T$), magnetic field lines, and velocity vectors,
in a main phase of the magnetic reconnection triggered by a supernova,
for the same model shown in figure \ref{Init} (Model A1).
A supernova appears at $(x,y)=(0,7)$, and disturbs the current sheet.
The tearing instability occurs, and
the current sheet gets thin ($t\sim 100-200$).
Sweet-Parker type reconnection occurs at $t\sim 200-250$.
Petschek type reconnection starts at $t\sim 270$.
The magnetic reconnection heats the gas,
and accelerates it by the magnetic tension force.
The hot gas of supernova flows into the diffusion region.
The magnetic islands are generated in the current sheet sequentially.
The heated gas is confined in the magnetic island.
\label{Temperature}}
\end{center}

\vspace{1mm}
\begin{center}
\caption{
Time evolution of 2D distribution of 
the gas pressure ($p_{\rm g}$), magnetic field lines, and velocity vectors,
in a main phase of the magnetic reconnection triggered by a supernova,
for the same model shown in figure \ref{Init} (Model A1).
The initial gas pressure is $p_{\rm g}=p_0+(p_0/\beta)\cosh^{-2}(y)$
where $p_0=0.6$ and $\beta=0.2$.
The gas pressure is high in the current sheet,
especially inside magnetic islands ($t\sim 270,300$).
\label{Pressure}}
\end{center}

\vspace{1mm}
\begin{center}
\caption{
Time evolution of 2D distribution of 
the current density ($J$), magnetic field lines, and velocity vectors,
in a main phase of the magnetic reconnection triggered by a supernova,
for the same model shown in figure \ref{Init} (Model A1).
The initial magnetic field is $B_{\rm x}=B_0\tanh(y)$ and $B_{\rm y}=0$
where $B_0=8.58$.
The plasma $\beta$ is $p_0/(B_0^2/8\pi)=0.2$.
The initial current-sheet thickness is $2l^{\rm init}=2$.
The initial current density is $\sim B_0/2\sim 4.5$ in the current sheet.
The current density increases in the diffusion region.
Petschek type reconnection generates two standing slow shocks
in the current sheet (see also figure \ref{SShock}).
The slow shocks appear also behind
the moving magnetic islands (see the figures at $t=270$ and 300). 
\label{Current}}
\end{center}

\end{figure}
\begin{figure}[p]

\vspace{1mm}
\begin{center}
\caption{The schematic illustration of the results of our simulation
(Model A1).
A supernova disturbs the current sheet (figure a).
The tearing instability occurs long after the passage of fast shock wave
(figure b).
The current-sheet thickness decreases.
Sweet-Parker type reconnection occurs (figure c).
The secondary tearing instability occurs in the Sweet-Parker current sheet
(figure d).
A magnetic island is generated (figure e).
Petschek type reconnection starts after the islands moves away (figure f).
It heats the interstellar gas, and accelerates the heated gas
by the magnetic tension force to the Alfv\'en velocity.
Many islands are generated sequentially,
and confine the heated gas.
\label{Scenario}}
\end{center}

\vspace{1mm}
\begin{center}
\caption{
Time variation of the half-thickness of the current sheet ($l$),
which is defined by the minimum half-width of half-maximum
of the gas pressure calculated parallel with $y$-axis in $|x|<25$,
for the same model shown in figure \ref{Init} (Model A1).
The supernova shock wave perturbs the current sheet ($t<2$).
The magnetic field dissipates slowly ($t<100$).
The tearing instability occurs ($t\sim 100-200$).
The current-sheet thinning occurs during the tearing instability.
Sweet-Parker type reconnection occurs ($t\sim 200-250$).
The secondary tearing instability occurs in the Sweet-Parker current sheet
($t\sim 250-270$).
The current-sheet thickness decreases again 
by the secondary tearing instability.
Petschek type reconnection occurs ($t\sim 270-400$). 
The current sheet gets much thin during the Petschek type reconnection.
At $t\sim 390-400$, the current sheet is very thin
because the magnetic island is moving.
\label{Thickness}}
\end{center}

\end{figure}
\begin{figure}[p]

\vspace{1mm}
\begin{center}
\caption{
The schematic illustration of
the time variation of the current-sheet thickness.
At first, the magnetic field dissipates slowly.
Second, the current-sheet thinning occurs
by the tearing instability in a nonlinear phase.
Sweet-Parker type reconnection occurs in the thin current sheet.
Next, the secondary tearing instability occurs in the Sweet-Parker current sheet.
The current-sheet thickness decreases more.
Petschek type reconnection starts.
The solid line moves to upper side for large-$\beta$, small-$y_{\rm ex}$, and 
large-$p_{\rm ex}$ models,
while it moves to lower side for small-$\beta$, large-$y_{\rm ex}$, and 
small-$p_{\rm ex}$ models.
\label{ThicknessPic}}
\end{center}

\vspace{1mm}
\begin{center}
\caption{
Time variation of the magnetic flux,
for the same model shown in figure \ref{Init} (Model A1).
The magnetic flux is defined by
$\psi (t)
=\int_0^t \max_{|x|<10}
|   (B_{\rm x} v_{\rm y})[\rm at ({\it x}, 2)]
+\it(B_{\rm x} v_{\rm y})[\rm at ({\it x},-3)]|$dt.
The magnetic flux increases during the tearing instability.
The magnetic flux grows by $\log\psi (t)\sim 0.00257t$,
which is shown by the dashed line.
The rate is 1/2 times the growth rate of the linear phase (\cite{mag99}).
\label{Flux}}
\end{center}

\vspace{1mm}
\begin{center}
\caption{
Time variation of the inflow velocity toward the diffusion region,
for the same model shown in figure \ref{Init} (Model A1).
Figure (a) shows the inflow velocity, defined by
$v_{\rm in}=\max_{|x|<10}
\left[     v_{\rm y}[\rm at ({\it x},-3)]
      -\it v_{\rm y}[\rm at ({\it x}, 2)] \right]/2$.
Figure (b) shows the Alfv\'en velocities at $(x_0,2)$ (dotted line)
and $(x_0,-3)$ (solid line),
where $x_0$ is the point where maximum inflow velocity is attained.
Figure (c) shows the reconnection rate defined by
$\epsilon=v_{\rm in}/v_{\rm A,s}$, 
where $v_{\rm A,s}$ is smaller one between
$v_{\rm A}[\rm at ({\it x_0},-2)]$ and $v_{\rm A}[\rm at ({\it x_0},2)]$.
The inflow velocity and reconnection rate are
$v_{\rm in}\sim 0.1-0.2$ and $\epsilon\sim 0.05-0.1$
during Petschek type reconnection,
while they are $\sim 0.05$ and $\sim 0.02$
during Sweet-Parker type reconnection.
At the beginning of Sweet-Parker type reconnection ($t\sim 200-220$),
they increases to $\sim 0.16$ and $\sim 0.6$, respectively,
because of the nonlinear growth of the tearing instability
(see the arrow).
\label{Inflow}}
\end{center}

\end{figure}
\begin{figure}[p]

\vspace{1mm}
\begin{center}
\caption{
Time variation of the various energies, showing
how the energy conversion occurs by the magnetic reconnection
in a central region ($|x|<25$ and $|y|<3$)
for the same model shown in figure \ref{Init} (Model A1).
The magnetic energy is converted to the thermal energy
slowly by Sweet-Parker type reconnection and secondary tearing instability
($t\sim 200-270$),
and quickly by Petschek magnetic reconnection ($t\sim 270-360$).
The magnetic islands suppress the magnetic reconnection
when they are staying at the reconnection region
($t\sim 310$ and 360).
The magnetic energy-release rate increases to $\sim 40$
at the beginning of Sweet-Parker reconnection
because the inflow velocity increases (see the arrow).
Maximum energy release rate is $-dE_{\rm mag}/dt\sim 70-80$ 
during Petschek type reconnection ($t\sim 340$).
The non-dimensional thermal energy of the initial supernova is 
$E_{\rm ex}\sim 1800$ in the numerical simulation.
Main thermal energy is supplyed from the magnetic energy,
not from the initial supernova.
\label{Energycen}}
\end{center}

\vspace{1mm}
\begin{center}
\caption{
The schematic illustration of the time variation of 
the magnetic energy-release rate in the reconnection region.
The magnetic dissipation occurs slowly at first.
The tearing instability starts much after the passage of supernova shock.
Sweet-Parker type reconnection occurs after current-sheet thinning
by the nonlinear phase of the tearing instability.
The magnetic energy-release rate increases once 
at $t\sim 200-220$ (see the arrow) because the inflow velocity increases.
After then, the magnetic energy is released slowly
by Sweet-Parker reconnection.
The secondary tearing instability occurs in Sweet-Parker current sheet.
The magnetic energy-release is suppressed by the small islands
generated by the secondary tearing instability.
Petschek type reconnection starts at last.
The magnetic energy is converted to the thermal energy violently
during Petschek type reconnection.
\label{EnergycenPic}}
\end{center}

\end{figure}
\begin{figure}[p]

\vspace{1mm}
\begin{center}
\caption{
Time variations of (a) the maximum drift velocity ($v_{\rm d}=J/\rho$),
(b) density $\rho$ where the maximum $v_{\rm d}$ is attained,
(c) current density ($J$) where the maximum $v_{\rm d}$ is attained,
and (d) maximum reconnection rate defined by $\eta J$
($-2.6<y<1, |x|<10$), 
for the same model shown in figure \ref{Init} (Model A1).
The current density increases to $J\sim B/l_{\rm SP}\sim 20$
after Sweet-Parker reconnection starts.
The density in the current sheet decreases
during Sweet-Parker type reconnection
and secondary tearing instability ($t\sim 200-270$).
The drift velocity exceeds the critical value ($v_{\rm c}=100$)
so that the anomalous resistivity sets in ($t\sim 270$).
Petschek type reconnection starts so that the reconnection rate increases.
\label{Drift}}
\end{center}

\vspace{1mm}
\begin{center}
\caption{
Profile of the variables in $x=-12$ at $t\sim 342$,
for the same model shown in figure \ref{Init} (Model A1).
$p_{\rm g}$, $T$, $\rho$, and $|v_{\rm y}|$
are higher in the current sheet than those outside the sheet.
The magnetic field strength is lower in the current sheet.
The profile of the current density has two peaks
($y\sim -2.3$ and $0.7$), corresponding to the slow shocks.
The density ratio across the shock front is $X=n_2/n_1\sim 0.3$,
where subscripts 1 and 2 mean that
the values are attained at $y\sim -2.1$ and $-0.3$, respectively.
The as pressure ratio is $p_{\rm g2}/p_{\rm g1}\sim 0.1$,
while 0.47 is required by Rankine-Hugoniot relation.
\label{SShock}}
\end{center}

\vspace{1mm}
\begin{center}
\caption{
2D distribution of the gas pressure at $t=160$ (a), 300 (b), and 450 (c),
for the same model shown in figure \ref{Init} (Model A1).
Figures (a), (b), and (c) show the distributions
during the tearing instability,
a little after Petschek type reconnection starts,
and after Petschek type reconnection stops.
The gas is ejected from the diffusion region
toward the left and right boundaries.
The gas pressure increases in the current sheet near the boundaries.
It extends the current sheet near the boundaries.
Furthermore, the magnetic tension suppresses the magnetic reconnection
because the magnetic field is bent outside the current sheet.
\label{Whole}}
\end{center}

\end{figure}
\begin{figure}[p]

\vspace{1mm}
\begin{center}
\caption{
Time variation of the magnetic energy-release rate ($|x|<25$, $|y|<3|$),
for Model B1 ($\beta=0.1$; dotted line), Model A2 ($\beta=0.2$; solid line),
and Model B2 ($\beta=0.3$; dashed line).
The magnetic energy-release rate decreases with $\beta$.
The maximum magnetic energy release rates are
$-dE_{\rm mag}/dt\sim 250$, $70$, and $20$, respectively.
\label{Emag_beta}}
\end{center}

\vspace{1mm}
\begin{center}
\caption{Dependence on $\beta$ of
(a) the maximum energy release rate ($dE_{\rm mag}/dt$)
in the reconnection region ($|x|<25$, $|y|<3$),
(b) the time when maximum $dE_{\rm mag}/dt$ is attained,
(c) maximum temperature of heated gas,
and (d) maximum velocity of the reconnection jet
(Model A2, B1-5, and E).
(a) The magnetic energy-release rate decreases with $\beta$.
The solid line and dotted line show
the maximum magnetic energy release rates
by Petschek and Sweet-parker type reconnections,
respectively.
The rates are determined by
Poynting flux entering into the reconnection region.
The theoretical $\beta$-dependence is shown by 
the dashed line
[$2(B_0^2/4\pi)S\epsilon v_{\rm A}^{\rm init}\propto\beta^{-3/2}$
where we assume $\epsilon\sim 0.1$ and $S\sim 26$]
for Petschek type reconnection,
and the dashed-dotted line 
[$2(B_0^2/4\pi)SR_{\rm m,SP}^{-1/2}v_{\rm A}^{\rm init}
\propto\beta^{-5/4}$
where we assume $R_{\rm m,SP}\sim 10^4$ and $S\sim 67$]
for Sweet-Parker type reconnection.
The results of the uniform-resistivity model (Model E)
are also shown by the triangle ($\triangle$).
(b) The time scale increases with $\beta$.
The magnetic phenomena occur in the Alfv\'en time
($\tau_{\rm A}\propto\beta^{1/2}$; dashed line)
or the time scale of tearing instability
($\tau_{\rm A}\propto\beta^{1/4}$; dashed-dotted line).
(c) The maximum temperature decreases with $\beta$.
It is determined by $T_{\rm max}\propto (1+1/\beta)$.
(d) The maximum velocity decreases with $\beta$.
The gas is accelerated to Alfv\'en  velocity
$v_{\rm A}\propto\tau_{\rm A}^{-1}\propto\beta^{-1/2}$.
\label{Beta}}
\end{center}

\end{figure}
\begin{figure}[p]

\vspace{1mm}
\begin{center}
\caption{Dependence of the results on the distance ($y_{\rm ex}$)
between the supernova and current sheet
(Models A2 and C1-3; see the caption of figure \ref{Beta}).
Figure (a) shows the maximum magnetic energy release rate
in the reconnection region ($|x|<25$, $|y|<3|$).
It decreases with $y_{\rm ex}$.
Figure (b) shows the time when maximum magnetic energy release rate is attained.
It increases with $y_{\rm ex}$ because the initial perturbation decreases.
Figures (c) and (d) show the maximum temperature of the heated gas
and maximum velocity of the reconnection jet.
They do not depend on $y_{\rm ex}$. 
\label{Distance}}
\end{center}

\vspace{1mm}
\begin{center}
\caption{
Dependence of the results on the energy of the initial explosion
($E_{\rm ex}\propto p_{\rm ex}\sim p_{\rm ratio}$)
(Models A2 and D1-4; see the caption of Figure \ref{Beta}).
Figure (a) shows the maximum magnetic energy release rate
in the reconnection region ($|x|<25$, $|y|<3$).
It does not depend on $p_{\rm pratio}$.
Figure (b) shows the time
when maximum magnetic energy release rate is attained.
It does not depend on $p_{\rm ratio}$.
Figure (c) shows the maximum temperature of the gas
heated by the magnetic reconnection.
It increases with $p_{\rm ratio}$
because the hot gas flows into the reconnection region.
Figure (d) shows the velocity of the reconnection jet.
It does not depend on $p_{\rm pratio}$
because the velocity of the reconnection jet is determined by Alfv\'en velocity.
The results of two-supernovae[at $(0,7)$ and $(0,-7)$]-model (Model D5)
are indicated by $\times$.
The magnetic reconnection starts early in the model
because the current sheet is compressed from both sides.
\label{Pratio}}
\end{center}

\vspace{1mm}
\begin{center}
\caption{
Time variations of the half-thickness of the current sheet ($l$),
defined by the minimum half-width of half-maximum of
the gas pressure in $|x|<25$,
for Model A2 ($\eta_0=0.015$; thick solid line),
Model F1 ($\eta_0=0.03$; thin dotted line),
Model F2 ($\eta_0=0.0075$; thin dashed-dotted line),
Model F3 ($\eta_0=0.00375$; thin dashed and double-dotted line),
Model F4 ($\eta_0=0.001$; thin dashed line),
Model F5 ($\eta_0=0.0001$; thin solid line),
and Model F6 ($\eta=0$; thick dashed line).
In Model A2 ($\eta_0=0.015$),
the tearing instability ($t\sim 60-150$) is followed by
Sweet-Parker type reconnection ($t\sim 150-200$),
secondary tearing instability ($t\sim 200-240$), 
and Petschek type reconnection ($t> 240$)
(figure \ref{Thickness}).
In zero-resistivity model (F6), the current sheet does not dissipates.
It starts to become thin at $t\sim 220$,
and Petschek type reconnection occurs by only numerical resistivity
(resistivity-like effect by the numerical noise).
In Model F1 ($\eta_0=0.03$), the current sheet dissipates early ($t<100$). 
\label{Thickness_eta}}
\end{center}

\end{figure}
\begin{figure}

\vspace{1mm}
\begin{center}
\caption{
Time variations of the magnetic energy-release rate
in the reconnection region ($|x|<25$, $|y|<3$),
for Model A1 ($\triangle x=0.20$; solid line)
and Model A2 ($\triangle x=0.25$; dotted line).
The magnetic energy-release rate does not depend on the grid size
because the ``background uniform resistivity'' ($\eta_0$) is
enough large to suppress the numerical resistivity.
The time scale of phenomena changes
because the grid number of the initial supernova changes,
and because the perturbation on the current sheet changes.
\label{Emag_grid}}
\end{center}

\vspace{1mm}
\begin{center}
\caption{
The schematic illustration of the time variation of
the current-sheet thickness in actual interstellar medium.
Here we assume $B\sim 10\ \mu$G, $n\sim 0.1$ cm$^{-3}$, $T\sim 10^4K$,
and Spitzer conductivity.
The current-sheet thickness decreases to the ion Lamor radius
($\sim 10^7$ cm)
by the nonlinear phase of secondary tearing instability
or by successive tearing instabilities.
When it decreases to the radius,
Petschek type reconnection can occur in the interstellar medium.
\label{Actual}}
\end{center}

\vspace{1mm}
\begin{center}
\caption{
2D distribution of the thermal radiation ($F\propto n^2T^a$)
for the same model shown in figure \ref{Init} (Model A1).
$a$ is $0.5$, $-0.6$, $2.9$, or $1.5$, respectively,
when the unit of temperature is assumed to be $T_0\sim 10^7$ K (figure [a]),
$\sim 10^{5-6}$ K (figure [b]), $10^4$ K (figure [c]), or $10^3$ K (figure [d]),
The high-temperature gas in the island and current sheet is bright.
\label{Xray}}
\end{center}

\end{figure}

\newpage
\scriptsize

\begin{center}
\begin{deluxetable}{lcrrrrrrrrrrr}
\tablecaption{\scriptsize
Parameters.  
We studied the effect of the grid size (Models A1-2),
plasma $\beta$ (Models B1-5),
distance $y_{\rm ex}$ between the current sheet and supernova (Models C1-3),
energy of initial supernova (Models D1-5),
resistivity model (Models E and F1-4),
and simulation region size (Models G1-2). Typical model is Model A1.
\label{Parameter}}
\tablewidth{0pt}
\tablehead{
\colhead{Model}  &
\colhead{$\beta$\tablenotemark{a}} &
\colhead{$y_{\rm ex}$\tablenotemark{b}} &
\colhead{$p_{\rm ratio}$\tablenotemark{c}} &
\colhead{$\eta_0$\tablenotemark{d}} &
\colhead{$\alpha$\tablenotemark{e}} &
\colhead{SN\tablenotemark{f}} &
\colhead{$\triangle x$\tablenotemark{g}} &
\colhead{$\triangle y$\tablenotemark{h}} &
\colhead{$L_{\rm x}$\tablenotemark{i}} &
\colhead{$L_{\rm y}$\tablenotemark{j}} &
\colhead{$N_{\rm x}$\tablenotemark{k}} &
\colhead{$N_{\rm y}$\tablenotemark{l}}
}
\startdata
A1 &0.20& 7.0& 500&0.01500&0.1&1&0.20&0.025&241.8&136.5&465&602\nl
A2 &0.20& 7.0& 500&0.01500&0.1&1&0.25&0.025&153.2&136.5&301&602\nl
B1 &0.10& 7.0& 500&0.01500&0.1&1&0.25&0.025&153.2&136.5&301&602\nl
B2 &0.15& 7.0& 500&0.01500&0.1&1&0.25&0.025&153.2&136.5&301&602\nl
B3 &0.30& 7.0& 500&0.01500&0.1&1&0.25&0.025&153.2&136.5&301&602\nl
B4 &0.50& 7.0& 500&0.01500&0.1&1&0.25&0.025&153.2&136.5&301&602\nl
B5 &1.00& 7.0& 500&0.01500&0.1&1&0.25&0.025&153.2&136.5&301&602\nl
C1 &0.20&10.0& 500&0.01500&0.1&1&0.25&0.025&153.2&136.5&301&602\nl
C2 &0.20&14.0& 500&0.01500&0.1&1&0.25&0.025&153.2&136.5&301&602\nl
C3 &0.20&28.0& 500&0.01500&0.1&1&0.25&0.025&153.2&136.5&301&602\nl
D1 &0.20& 7.0&  50&0.01500&0.1&1&0.25&0.025&153.2&136.5&301&602\nl
D2 &0.20& 7.0& 100&0.01500&0.1&1&0.25&0.025&153.2&136.5&301&602\nl
D3 &0.20& 7.0& 250&0.01500&0.1&1&0.25&0.025&153.2&136.5&301&602\nl
D4 &0.20& 7.0&1000&0.01500&0.1&1&0.25&0.025&153.2&136.5&301&602\nl
D5 &0.20&$\pm$7.0& 500&0.01500&0.1&2&0.25&0.025&153.2&136.5&301&602\nl
E  &0.20& 7.0& 500&0.01500&0.0&1&0.25&0.025&153.2&136.5&301&602\nl
F1 &0.20& 7.0& 500&0.03000&0.0&1&0.25&0.025&153.2&136.5&301&602\nl
F2 &0.20& 7.0& 500&0.00750&0.1&1&0.25&0.025&153.2&136.5&301&602\nl
F3 &0.20& 7.0& 500&0.00375&0.1&1&0.25&0.025&153.2&136.5&301&602\nl
F4 &0.20& 7.0& 500&0.00100&0.1&1&0.25&0.025&153.2&136.5&301&602\nl
F5 &0.20& 7.0& 500&0.00010&0.1&1&0.25&0.025&153.2&136.5&301&602\nl
F6 &0.20& 7.0& 500&0.00000&0.0&1&0.25&0.025&153.2&136.5&301&602\nl
G1 &0.50& 7.0& 500&0.01500&0.1&1&0.25&0.025&153.2& 66.7&301&349\nl
G2 &0.50& 7.0& 500&0.01500&0.1&1&0.25&0.025&153.2& 44.9&301&270\nl
\enddata
\tablenotetext{a}
{the ratio of the gas to magnetic pressure outside the current sheet}
\tablenotetext{b}{the distance between the supernova and current sheet}
\tablenotetext{c}{the initial gas pressure of the supernova}
\tablenotetext{d}{``background uniform resistivity''}
\tablenotetext{e}{the parameter of anomalous resistivity;
0.1 for the anomalous resistivity, and 0.0 for the uniform resistivity}
\tablenotetext{f}{number of the initial supernovae}
\tablenotetext{g}{minimum grid size in $x$-axis}
\tablenotetext{h}{minimum grid size in $y$-axis}
\tablenotetext{i}{simulation region size in $x$-axis}
\tablenotetext{j}{simulation region size in $y$-axis}
\tablenotetext{k}{number of grid points in $x$-axis}
\tablenotetext{l}{number of grid points in $y$-axis}
\end{deluxetable}
\end{center}

\end{document}